\documentclass[12pt,oneside]{article}
\usepackage{amsfonts,amssymb,graphicx}
\def\be{\begin{equation}}
\def\ee{\end{equation}}
\def\bea{\begin{eqnarray}}
\def\eea{\end{eqnarray}}

\def\<{\langle}
\def\>{\rangle}
\def\~{\tilde}
\def\s{\sigma}
\def\l{\lambda}
\def\L{\Lambda}

\def\o{\omega}

\newcommand{\Z}{\mathbb Z}

\newcommand{\av}[1]{\mbox{{\rm Av}}\left(#1\right)}

\newtheorem{theorem}{Theorem}

\newtheorem{corollary}{Corollary}

\newenvironment{proof}{Proof:}{\hfill$\square$\vskip.5cm}
\newenvironment{proofs}{proof:}{\hfill$\square$\vskip.5cm}
\begin{document}
\begin{center}
\vspace{1truecm}
{\bf\sc\Large correlation inequalities for spin glasses}\\
\vspace{1cm}
{Pierluigi Contucci $^{\dagger}$,\quad Joel Lebowitz $^{\ddagger}$}\\
\vspace{.5cm}
{\small $^\dagger$ Dipartimento di Matematica, Universit\`a di Bologna, {e-mail: {\em contucci@dm.unibo.it}}} \\
\vspace{.5cm}
{\small $^\ddagger$ Department of Mathematics, Rutgers University, {e-mail: {\em lebowitz@math.rutgers.edu}}}\\
\end{center}
\vskip 1truecm
\begin{abstract}\noindent
We prove a correlation type inequality for spin systems with quenched symmetric
random interactions.  This gives monotonicity of the pressure
with respect to the strength of the interaction for a class of spin
glass models. Consequences include 
existence of the thermodynamic limit for the pressure and bounds on
the surface pressure.  We also describe other conjectured inequalities for such systems.
\end{abstract}
\newpage\noindent
\section{Introduction}
Correlation inequalities have played and continue to play an important
role in many areas of 
statistical mechanics. In addition to describing microscopic structure
they also provide information about macroscopic properties: for 
ferromagnetic spin systems they give monotonicity of the critical temperature, inequalities
for critical exponents, etc.

It would clearly be desirable to obtain similar results also for disordered
systems, like spin glasses, in which both ferromagnetic and anti-ferromagnetic
interactions are present with assigned
probabilities.  In this note we obtain some correlation type 
inequalities for such systems. These yield various results usually 
obtained for ferromagnetic systems from the first GKS inequalities (\cite{Gr,Gr2,KS}), e.g. monotonicity 
and hence existence of the thermodynamic limit of the pressure, and bounds  
on the surface pressure. We also discuss intriguing examples of versions of the second type GKS 
inequalities, but at the moment a general proof is still lacking.

\section{Definitions and Results}

Let 
$\s_n=\pm 1$, $n\in \Lambda\subset \Z^d$, and denote by 
$\Sigma_\Lambda$ the set of all $\s=\{\s_n\}_{n\in \Lambda}$, and
$|\Sigma_\Lambda|=2^{|\Lambda|}$. To each  $X\subset\Lambda$ 
we associate the spin product $\s_X=\prod_{i\, \in X}\s_i$ and
random variable $J_X$. The $J$'s, denoted collectively by ${\bf J}$, are a family of 
independent unit symmetric random variables: 
\be\label{symme}
\av{J_X}=0 \; ,
\ee
\be\label{indy}
\av{J_XJ_Y}=\delta_{X,Y} \; ,
\ee
where the $\av{}$ denotes the average over the distributions of the $J$'s.
The energy of a configuration
$\sigma$ is given by 
\be
U_{\Lambda}(\s) \; = \; -\sum_{X\subset \Lambda} \lambda_X J_X\s_X +
\bar U_{\Lambda}(\sigma) \; ,
\label{hami}
\ee
where $\bar U_{\Lambda}$ is a non-random (reference) interaction which we need
not specify in any detail for the moment.  It can in particular be set
equal to zero.The $\{\lambda_X\}$, denoted
collectively by ${\bf \lambda}$, are auxiliary parameters which are used
to tune the strength of the different interactions. Some can be set
(at the end) equal to zero.

The probability distribution for specified $\bf{\lambda}$ and $\bf J$
is given by
$\mu = Z_\Lambda^{-1}e^{-U_\Lambda}$ with random partition function
\be 
Z_\Lambda \; = \; \sum_{\s} e^{-U_{\Lambda}(\s)}.
\ee 
The quenched pressure is given by 
\be P_\Lambda \; = \;
\av{\log Z_\Lambda} \; .  
\ee 
We define the thermal correlations for a given
$\bf J$ by 
\be 
\o_A \; = \; \frac{\sum_{\s} \s_A 
e^{-U_{\Lambda}(\s)} } {\sum_{\s} e^{-U_{\Lambda}(\s)}} \; .  
\ee 
The
quenched correlations $\bar \omega_A$ are given by $\av{\omega_A}$ and
we also define (dropping the subscript $\Lambda$) 
\be 
\rho_A \; = \; \av{J_A \o_A} = \frac{\partial
P}{\partial \lambda_A} \; , 
\ee 
The pressure (and the $\bar \omega_A$) is clearly an even function of each $\lambda_X$ 
and hence $\rho_A$ is an odd function of $\lambda_A$ and an 
even function of $\lambda_X$, $X \ne A$.
\begin{theorem}\label{teo1}
Let $\lambda_A 
\geq 0$, then $\rho_A \geq 0$. Consequently the
 quenched pressure  
 $P$ is monotone increasing with respect to each $\lambda_A$. i.e.
\be\label{mono}
\frac{\partial P}{\partial \lambda_A} \; = \; \rho_A \ge 0, \quad \forall  A \quad with \quad  
 \lambda_A \geq 0 \; .
\ee
\end{theorem}
\noindent {\bf Remark}
Theorem 1 immediately implies that 
 the pressure $P({\bf \lambda})$ has a minimum for $\lambda_A = 0$, for any
 $A \in \Lambda$.  E.g.\ 
the pressure of a system with symmetric 
nearest-neighbor random interactions is smaller than the one with added next nearest
neighbor interactions. Another consequence is that in the case when
 $\bar{U}_\Lambda=0$ or containing only one site potentials, e.g.\ for standard 
spin glasses, the pressure for periodic boundary
conditions is larger than the one with free boundary conditions.

\begin{proof}
The function $P$ is symmetric and convex with respect to $\lambda_A$,
i.e. 
\be\label{symm}
\rho_A=\frac{\partial P}{\partial \lambda_A} = 0  \quad {\rm for}~~ \lambda_A = 0.
\ee
and 
\be\label{conv}
\frac{\partial^2 P}{\partial \lambda^2_A} \; = \; 
\av{J^2_{A}[1-\o^2_A] } \; \ge 0 \; .
\ee
The 
first equality follows from symmetry and the second from the fact that
$\omega^2_A \leq 1$.  Consequently from (\ref{conv}) 
\be
\rho_A(\lambda^{(1)}_A)\ge \rho_A(\lambda^{(2)}_A) \; ,  \quad \forall \; \lambda_A^{(1)} 
\geq \lambda_A^{(2)} \geq  \; 0
\ee
and the theorem follows by chosing $\lambda^{(2)}_A=0$ and from (\ref{symm}).
\end{proof}

\begin{theorem}
\label{teo2}
When $\bar U_\Lambda$ is just a sum of one site potentials, i.e.\
$\bar U_\Lambda = \sum_{i \in \Lambda} h_i \sigma_i$,
the pressure $P$ is super-additive, i.e., 
for all disjoint decompositions $\Lambda=\cup_{h=1}^{K} \Lambda_h$, 
$\Lambda_l\cap \Lambda_m=\emptyset$ $l\neq m$
\be\label{sa}
P_\L \ge \sum_{h=1}^{K} P_{\L_h} \; .
\ee
\end{theorem}
{\bf Remark}  These properties of $P$ are 
similar to those for the ferromagnetic case
\cite{Gr2} with the $\rho_A \geq 0$ playing the role of the first GKS
inequality. This result generalizes the one obtained for the Gaussian case in \cite{CG}.
\begin{corollary}
Assuming thermodynamic stability, 
\be
\av{e^{-U}} \; \le \; e^{c|\Lambda|} \; ,
\ee
the super-additivity provided by Theorem 2 implies, by standard methods,
monotonicity and existence of the thermodynamic limit for the
pressure with free boundary conditions.  In particular
\be
p \; = \; \lim_{\Lambda\nearrow\Z^d}\frac{P_\Lambda}{|\Lambda|} \; = \; 
\sup_{\Lambda}\frac{P_\Lambda}{|\Lambda|}\; .
\ee
\end{corollary}
{\bf Remark} Decomposing a $d$-dimensional hypercube $\Lambda$ into $(d-1)$-dimensional 
hypercubes  $(\ref{sa})$ immediately implies that 
the pressure in dimension $d$, $p^{(d)}$ is an increasing function of $d$
\be
p^{(d+1)} \; \ge \; p^{(d)}
\ee
Corollary $(2.1)$ admits the following \begin{proofs} 
to each partition of $\Lambda$ we associate the interpolating potential for $0\le t\le 1$
\be
U_\Lambda(t) \, = \, \sum_{h=0}^{K}
t_hU_{\Lambda_h}^{(h)}, \quad \Lambda_0 = \L \, ,
\ee
with $t_0=t$ and $t_h=(1-t)$ for $1\le h\le K$,
\be
U_{\Lambda_h}^{(h)} \, = \, -\sum_{X\subset\Lambda_h} \l_XJ^{(h)}_X\s_X
\label{local}
\ee
and we have dropped the argument $\sigma$ and set $\bar U_n = 0$, for
notational convenience.
The $J^{(h)}_X$ are centered independent unit random variables
\be
\av{J^{(l)}_XJ^{(m)}_Y}=\delta_{l,m}\delta_{X,Y} \; .
\ee
We define the interpolating partition function
\be
Z_\Lambda(t) \, = \, \sum_{\s\in\Sigma_N}e^{-U_\Lambda(t)} \, ,
\ee
and we observe that
\be
\label{ifa}
Z_\Lambda(0) \, = \, \prod_{h=1}^{K}Z_{\Lambda_h}(J^{(h)}) \; , \; \quad
Z_\Lambda(1)=Z_{\Lambda}(J)\; .
\label{st}
\ee
We consider the interpolating pressure
\be 
P_\Lambda(t) := \, \av{\ln Z_\Lambda(t)} \; .
\label{fe}
\ee
Using  (\ref{st}) we get
\be
\label{ifa2}
P_\Lambda(0) \, = \, \sum_{h=1}^{K}P_{\Lambda_h} \; , \; \quad
P_\Lambda(1)=P_{\Lambda} \; .
\ee
Considering ${\cal C}_\Lambda$ the set of all subsets of $\L$ with non empty intersection with more than 
one $\L_h$ a straightforward computation gives
\be\label{derif} 
\frac{d}{dt} P_\Lambda(t) \,=\,\sum_{X\in {\cal C}_\Lambda}\lambda_X^2\rho_X \ge 0
\; , 
\ee 
where the last inequality comes from theorem 1. Hence (\ref{ifa}) and (\ref{derif}), 
imply (\ref{sa}).
\end{proofs}
\begin{theorem}
For finite range interactions, i.e. $\lambda_X=0$ for $|X|\ge r$ (e.g.\ the nearest neighbor case)
the first correction to the leading term of the pressure, $T_\L$ defined by:
\be\label{sp}
P_\Lambda = p|\L|+T_\L
\ee
has a non positive value and is of surface size:
\be\label{sz}
c |\partial\L| \le T_\L \le 0 .
\ee
\end{theorem}
{\bf Remark} The result (\ref{sz}) is analogous to that for the ferromagnetic case in \cite{FL,FC}.
The theorem generalizes to all symmetric distributions of the $J_X$ the result obtained for the Gaussian
case in \cite{CG2}.

\begin{proof}
We give the proof for nearest-neighbor interactions, $r=2$, 
$U_{\Lambda}(\s) = \sum_{b} \lambda_b J_b\s_b$ where $b$ represents a
bond of the lattice $b=(n,n')$, $|n-n'|=1$. The extension to
the general case is straightforward and only changes the value of the constant
of (\ref{sz}). Given the $d$-dimensional cube $\Lambda$, $|\Lambda|=L^d$  consider  its
magnification $|\Lambda_k| = (kL)^d$. Clearly $\Lambda_k$ can be partitioned into $k^d$ disjoint cubes
$\Lambda_s$ all congruent to $\Lambda$. We call ${\cal C}_\Lambda$ the set of 
bonds connecting the different $\Lambda_s$. In finite volume and with free boundary
conditions we have by definition 
\be
P_\Lambda \; = \; \av{\ln Z_\Lambda} \; = \; k^{-d}\av{\ln
Z_\Lambda^{k^d}}\; .
\label{rep}
\ee
Since the limiting pressure per particle is independent of the 
boundary conditions we have, indicating by $\Pi$ the periodic boundary conditions (see \cite{FL})
\be
p|\Lambda| \; = \; \lim_{k\to\infty} k^{-d} \av{\ln Z^{(\Pi)}_{k\Lambda}} \;
\label{per}
\ee
By (\ref{rep}) and (\ref{per}) we obtain
\bea\nonumber
T_{\Lambda} \, &=&  \left( P_\Lambda - p|\Lambda| \right) \\
&=& \, \lim_{k\to \infty} k^{-d}\av{\ln Z_\Lambda^{k^d}- \ln
Z^{(\Pi)}_{k\Lambda}} \; .
\label{bs}
\eea
We chose now the $\lambda$'s in such a way that 
\be
\lambda_b \; = \; \left\{
\begin{array}{ll} 
\lambda, & \mbox{if $b\in {\cal C}_{\Lambda}$}, \\
1, & \mbox{otherwise} \; .
\end{array}\right.
\ee
(\ref{bs}) then becomes, using the fundamental theorem of calculus, 
\be
T_{\Lambda} \, = \, \lim_{k\to \infty}
k^{-d}\left[P^{(\Pi)}_\Lambda(0)-P^{(\Pi)}_\Lambda(1)\right] \, = \,
- \lim_{k\to \infty} k^{-d} \int_{0}^{1}\frac{d}{d\lambda}P^{(\Pi)}_\Lambda (\lambda)
d\l \; .
\label{bs2}
\ee
By a simple computation
\be
\label{derif2} 
\frac{d}{d\l}P^{(\Pi)}_\Lambda (\l) \; = \; \sum_{b\in{\cal C}_{\Lambda}} \rho^{(\Pi)}_b(\l) \; = \; |{\cal C}_{\Lambda}|
\rho^{(\Pi)}_b(\l)
\; ,
\ee 
where we have used the translation symmetry over the torus. By the identity 
\be
2|{\cal C}_\Lambda|=k^d|\partial\Lambda|
\ee
we obtain
\be
T_{\Lambda} = -  \frac{|\partial\Lambda|}{2} \lim_{k\to
\infty}\int_{0}^{1} \rho^{(\Pi)}_b(\l)d\l \; ,
\ee
which immediately entails the theorem.
\end{proof}
\section{Other Inequalities}
We have seen that Theorem 1 expresses for spin glasses a monotonicity property
that entails the same consequences that the first GKS inequality does for ferromagnetic
systems. From this perspective it is interesting to examine whether the quantity
\be\label{second}
\frac{\partial^2 P}{\partial \l_A \partial \l_B} \; = \; \av{J_AJ_B[\o_{AB}-\o_A\o_B]}
\ee
has a definite sign when $A\neq B$ (for $A=B$, $\omega_{AB} = 1$ and
the sign is positive). We have some evidence based on explicit
calculations (work in progress with F.Unguendoli) that for $A\neq B$ with the $J$'s satisfying (\ref{indy}), 
the sign is negative, but are unable at this time 
to produce a general proof or find a counter example. A simple computation 
based on integration by parts shows that such an inequality would imply
that the overlap expectations in Gaussian spin glasses would be monotonic 
non decreasing with the volume, similar to spin expectations in ferromagnetic systems.

Another interesting question is the case where the $J$'s are not
centered random variables but with $\av{J_A}=\mu_A >0$.
We then ask whether $\av{\omega_A} \geq 0$. This is essentially the same question as whether
$\av{\omega_A} \geq 0$ when $\bar U_\Lambda$ is ferromagnetic.  In
that case $\omega_A \geq 0$ when $\Lambda_X = 0$ for all $X \subset
\Lambda$ and it seems very reasonable to expect that this will be
preserved when we add symmetric random interactions. Also in this case
we have evidence that the inequality hold for specific cases (work in progress with F.Unguendoli) 
but unfortunately we don't have a general proof or a counter example. It is interesting to note that in a
subspace of the random parameter space in which variance and mean
of the $J$'s are chosen to be identical (Nishimori line) a suitable version of the GKS inequalities
can be proved with consequences similar to those
obtained here for the surface pressure (see \cite{MNC,CMN}).
\vskip .5truecm

{\bf Acknowledgments}. P.C. thanks  A. Van Enter, C.Giardina, S.Graffi, 
F.Guerra, F.Den Hollander, H.Nishimori, S.Morita and F.Unguendoli for useful discussions.
Research supported in part by NSF Grant DMR0442066 and by AFOSR Grant AF-FA99550-04.

\end{document}